\documentclass[8pt,floatfix,twocolumn,showpacs,
preprintnumbers,amsmath,amssymb,superscriptaddress,prl]{revtex4}

\usepackage{graphicx}
\usepackage{dcolumn}

\newcommand{\Eu}{Eu$_8$Ga$_{16}$Ge$_{30}$}
\newcommand{\Sr}{Sr$_8$Ga$_{16}$Ge$_{30}$}
\newcommand{\Ba}{Ba$_8$Ga$_{16}$Ge$_{30}$}
\newcommand{\X}{X$_8$Ga$_{16}$Ge$_{30}$}

\begin{document}

\title{Four-well tunneling states and elastic response of clathrates}
\author{I. Zerec}
\affiliation{Max-Planck-Institute for Chemical Physics of Solids,
D-01187 Dresden, Germany}
\author{V. Keppens}
\affiliation{Dept. of Materials Science and Engineering, The
  University of Tennessee, Knoxville, TN 37996}
\author{M. A. McGuire}
\affiliation{Dept. of Physics and Astronomy, The 
University of Mississippi, University, MS 38677}
\altaffiliation{Present address: Dept. of Physics, Cornell University,
  Ithaca, NY 14853}
\author{D. Mandrus}
\author{B. C. Sales}
\affiliation{Condensed Matter Sciences Division, Oak Ridge National
  Laboratory, P.O. Box 2008, Oak Ridge, TN 37831} 
\author{P. Thalmeier}
\affiliation{Max-Planck-Institute for Chemical Physics of Solids,
D-01187 Dresden, Germany}
\date{\today}

\begin{abstract}
We present resonant ultrasound elastic constant measurements of
the clathrate compounds Eu$_{8}$Ga$_{16}$Ge$_{30}$ and
Sr$_{8}$Ga$_{16}$Ge$_{30}$.  The elastic response of the Eu
clathrate provides clear evidence for the existence of a new type
of four-well tunneling states, described by two nearly degenerate
four level systems (FLS). The FLS's are closely linked with the
fourfold split positions of Eu known from neutron diffraction
density profiles. Using a realistic potential we estimate the
tunneling frequencies and show that the energy gap between the two
FLS's is of the same order as the Einstein oscillator frequency.
This explains why the observed harmonic oscillator type specific
heat is not modified by tunneling states. In addition the quadrupolar
interaction of FLS's with elastic strains explains the pronounced
depression observed in elastic constant measurements. In the case of
the Sr clathrate, we show that the shallow dip in the elastic
constant $c_{44}$ is explained using the same type of quadrupolar
interaction with a soft Einstein mode instead of a FLS.
\end{abstract}


\maketitle

Clathrate compounds are composed of polyhedral cages with 12-16
faces formed by Si, Ge or Ga atoms~\cite{Nolas01}. The oversized
cages accommodate guest atoms like Ba, Sr, or Eu which exhibit
large amplitude anharmonic motion. This leads to a strong
scattering of acoustic phonons and a small thermal conductivity.
In many cases the electrical conductivity remains high and the
resulting large thermoelectric figure of merit has led to
sustained interest in this class of compounds. The glass-like low
temperature thermal conductivity \cite{Cohn99,Paschen01} and
ultrasound attenuation \cite{Keppens00,Keppens02} have been
interpreted as evidence for the existence of tunneling states.
These are commonly discussed within a generic two well potential
model leading to two level system (TLS). The thermal conductivity
is more glass-like for clathrates with larger atomic displacement
parameters (ADP) of the guest atoms. For the clathrates discussed
here the ADP's are very large. The nuclear density map for Eu(2) in
\Eu{} clearly shows the distinct symmetrically
related four maxima, away from the cage
center~\cite{Sales01,Chakoumakos01}, leading to the fourfold split
site model. Large static ADP for Sr(2), and considerably smaller for
Ba(2), in \Sr{} and \Ba, respectively, point also to the split site
model~\cite{Sales01,Chakoumakos00,Paschen01}, although the nuclear
density maps do not show distinct fourfold maxima.

In this letter we present elastic constant measurements on \X{} (X =
Eu, Sr) that show anomalies that cannot be naturally explained
within a generic TLS model. Instead we propose two new models that
we feel are physically appealing because they draw their
inspiration from the nuclear density map results.

For \Eu{}  we introduce a four-well potential model with the proper
symmetry of split sites that describes a rotor type tunneling
motion. We show that for appropriate potential parameters two
partially degenerate FLS's exist. This model explains both the
specific heat and elastic anomalies and is compatible with neutron
diffraction.

For \Sr{}  the elastic response is explained using the same
quadrupolar interaction with elastic strains, but with local Sr
vibrations. This model may be also considered in other compounds
with soft local modes, such as skutterudites, where a similar
shallow dip in the elastic constant is observed~\cite{Keppens98},
without pronounced feature in the specific heat.\\

The elastic constants reported in this work were measured using
Resonant Ultrasound Spectroscopy (RUS) \cite{Migliori93}. RUS is a
novel technique for determining the elastic moduli of solids,
based on the measurement of the resonances of a freely vibrating
body:  in a RUS experiment, the mechanical resonances of a freely
vibrating solid of known shape are measured, and an iteration
procedure is used to 'match' the measured  lines with the
calculated spectrum.  This allows determination of all elastic
constants of the solid from a single frequency scan, eliminating
the need for separate measurements to probe different moduli. RUS
measurements were taken as a function of temperature using a
homemade probe that fits in a Physical Property Measurement System
(PPMS) from Quantum Design. Heat capacity measurements were also
performed in the PPMS using the apparatus from Quantum Design.

The four well potential is modeled, in accord with nuclear density
data, as
\begin{eqnarray}
\label{pot}
V(\rho,\phi)=\frac{V_0}{\rho^2}(1+\cos(4\phi))
+\frac{V_1}{\rho^2}+\frac{1}{2}K\rho^2.
\end{eqnarray}
The harmonic oscillator (HO) potential of the last term is
modified by the four-well potential term. V$_{0}$ is the potential
barrier in the azimuthal direction. The V$_{1}$ term determines
the positions of the minima, located in the directions
$\phi=(2n+1)\pi/4$, for $n=\{0,1,2,3\}$, at
$\rho_{min}=\sqrt[4]{2V_{1}/K}$. It also represents the infinite
potential barrier across the center of the cage. This means that
there is no tunneling through (and no nuclear density in) the
center, consistent with the nuclear density map for
Eu$_{8}$Ga$_{16}$Ge$_{30}$~\cite{Sales01,Chakoumakos01}. The HO
force constant is given by Einstein temperature, $\Theta$, of the
soft local mode of the guest atom of mass M,
$K=M\left(k_{B}\Theta/\hbar\right)^{2}$. $\Theta$ may be
determined from ADP, Raman, or specific heat data. The wave
function may be written, after separation of the variables, as the
product of angular and radial parts,
$\psi(\rho,\phi)_{n,m}=R(\rho)_{n,m}\Phi(\phi)_{m}$. The former
obeys
\begin{eqnarray}
\label{mu} \frac{\partial^2}{\partial\phi^2}\Phi_m
+(\mu_m-\bar{V}_0\cos(4\phi))\Phi_m=0,
\end{eqnarray}
%%%%%%%%%%%%%%%%%%%%%%%%%%%%%%%%%%%%%%%%%%%%%%%%%%%%%%%%%%%%%%%%%%%%%%%%%%%%
\begin{figure}
  \begin{center}
    \includegraphics[width=0.16175\textwidth]{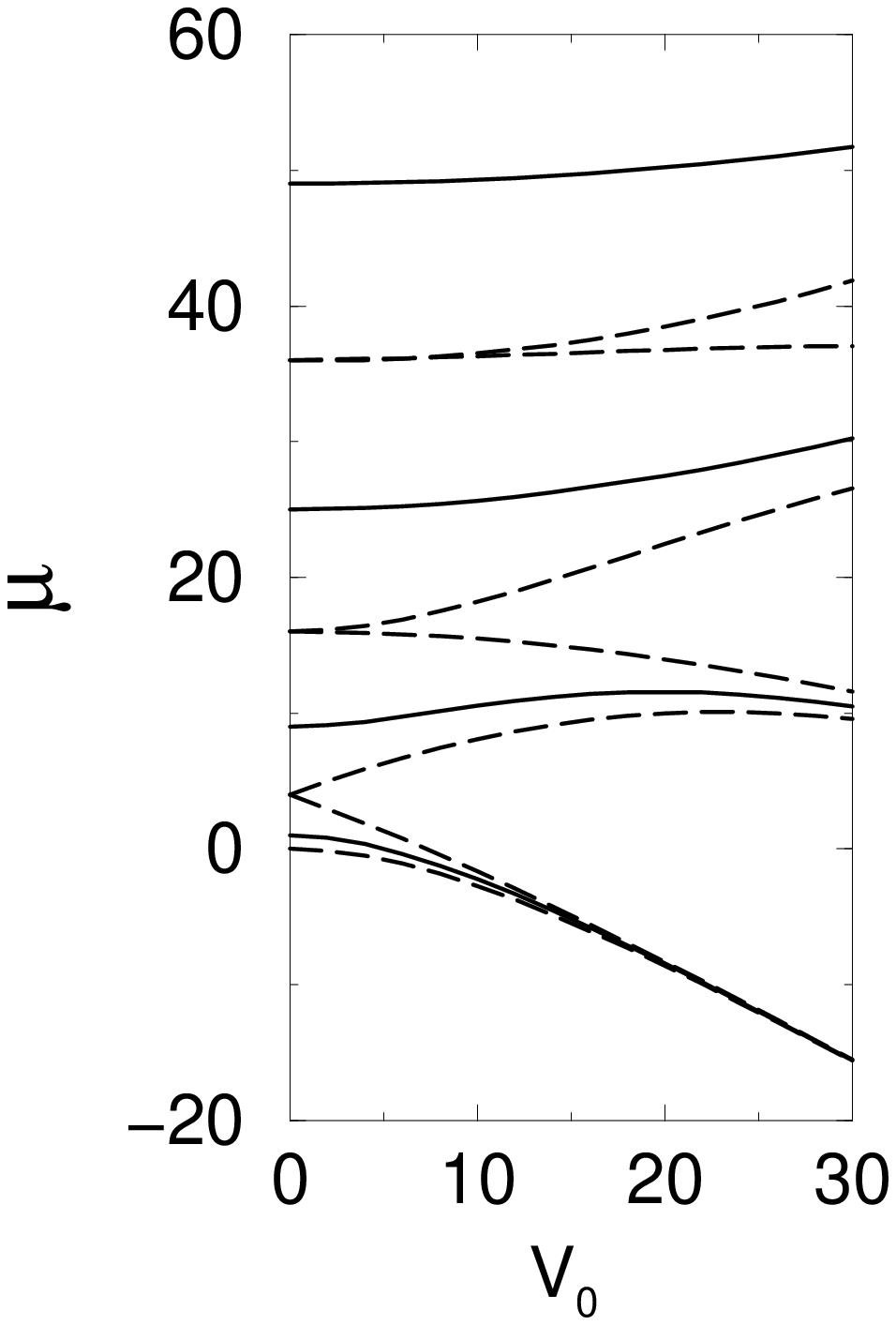}\hspace{0cm}
    \includegraphics[width=0.3\textwidth]{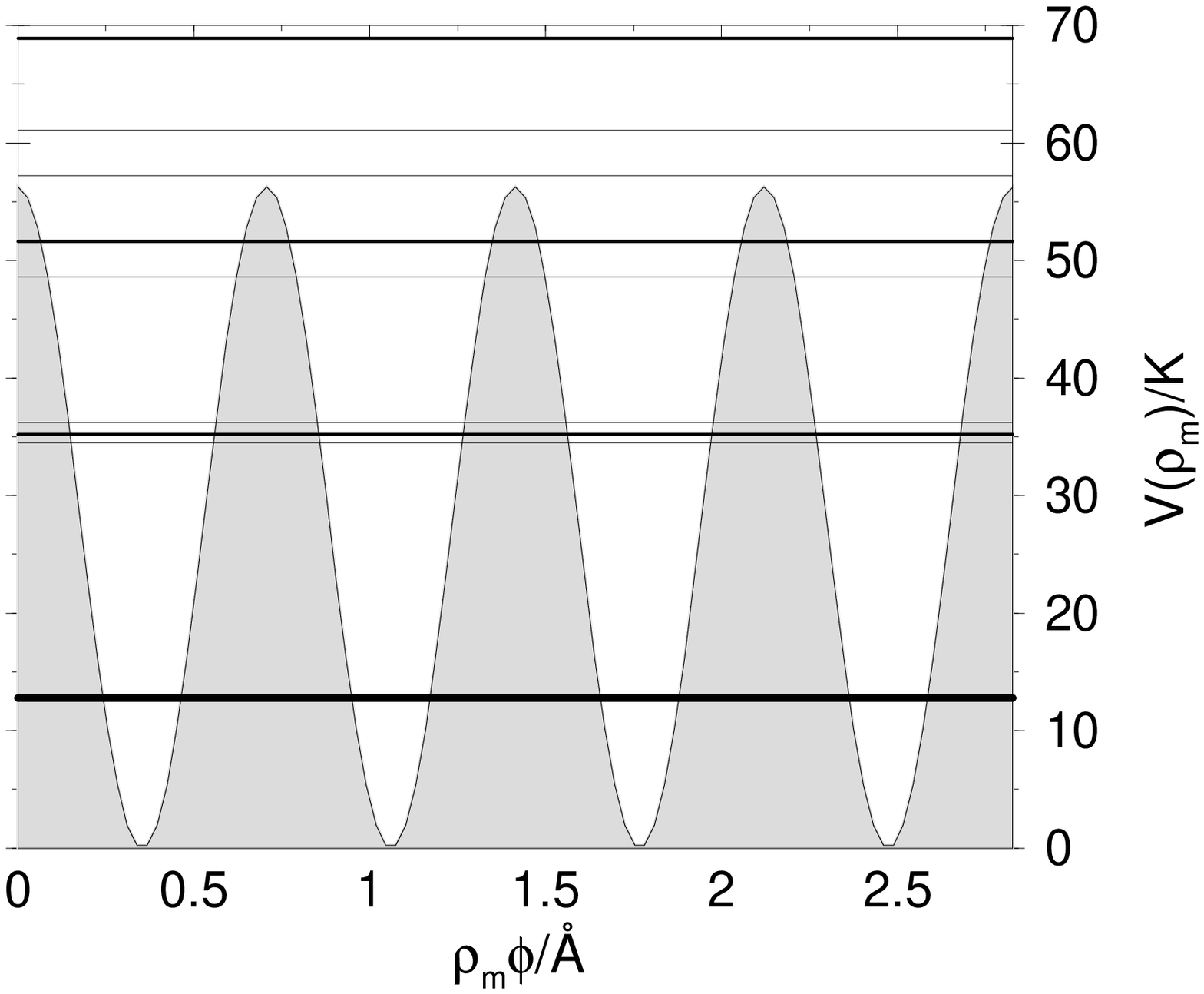}
    \caption{\textit{Left}: evolution of $\mu$ with
      potential barrier $\bar{V}_0$. Dashed lines represent nondegenerate 
      (for $\bar{V}_0 >0$) and the solid lines doubly degenerate
      levels. The formation of lower
      FLS$_{1}$ and upper FLS$_{2}$, may be followed.
      \textit{Right}: eigenvalues $\lambda_{nm}$ and potential V($\phi$,$\rho$)
      on the circle $\rho=\rho_{0}$ with parameters appropriate for \Eu. Lower
      FLS$_{1}$: fat line, doublets: thick line, nondegenerate state:
      thin line. Energies in units of K. For every eigenvalue
      $\mu_{m}$, there is a sequence of equidistant eigenvalues $\lambda_{n,m}$
      of the radial equation (Eq.\ref{partialR}) separated by
      $4\times\Theta/2=60$~K.\label{eigenmu}}
  \end{center}
\end{figure}
%%%%%%%%%%%%%%%%%%%%%%%%%%%%%%%%%%%%%%%%%%%%%%%%%%%%%%%%%%%%%%%%%%%%%%%%%%%%%
where $\bar{V_{0}}=\frac{2M}{\hbar^{2}}V_{0}$ is dimensionless.
For $\bar{V}_0=0$ eigenvalues $\mu_{m}^{(0)}=m^{2}$,
eigenfunctions $\Phi^{\pm}_{m}(\phi)=\cos m\phi$, $\sin m\phi$ and
angular momentum $m\hbar$ correspond to the free rotor.
This has a formal analogy to directional rotors formed by guest molecules
in clathrates \cite{Wuerger02}.
Introduction of $\bar{V}_0$ breaks the full azimuthal symmetry to
fourfold symmetry, lifts the twofold degeneracy of states with even
values of $m$ and leads to the formation of FLS with increasing
$\bar{V}_0$, shown in Fig.~\ref{eigenmu}. The radial wave function satisfies

\begin{eqnarray*}
\label{partialR}
\frac{1}{x}\frac{\partial}{\partial x}x\frac{\partial R_{n,m}}{\partial x}
+\left(\lambda_{n,m}-\frac{\mu_m+\bar{V}_0+\bar{V}_1}{x^2}-x^2
\right)R_{n,m} = 0.
\end{eqnarray*}

Here, $x^2=(M\,K/\hbar^2)^{1/2}\rho^2$ is the dimensionless radial
coordinate and $\lambda=2 E/(k_B\Theta)$ denotes dimensionless
eigenvalues

\begin{eqnarray}
\label{lambda0}
\lambda_{n,m}=2\left(\sqrt{\mu_m+\bar{V}_0+\bar{V}_1}+1\right)+4n.
\end{eqnarray}

The maxima of the radial wave functions correspond to the nuclear
density maxima. The tunneling splitting $\Delta_{1}$ in units of
Kelvin for the lowest FLS$_{1}$ is given by 
 
\begin{eqnarray}
\label{delta}
\Delta_1 \approx \frac{\hbar^2}{2 M k_B \rho_0^2}
\left(\mu_1(\bar{V}_0)-\mu_0(\bar{V}_0)\right).
\end{eqnarray}

\begin{figure}
  \begin{minipage}{5cm}
    \includegraphics[width=.9\textwidth]{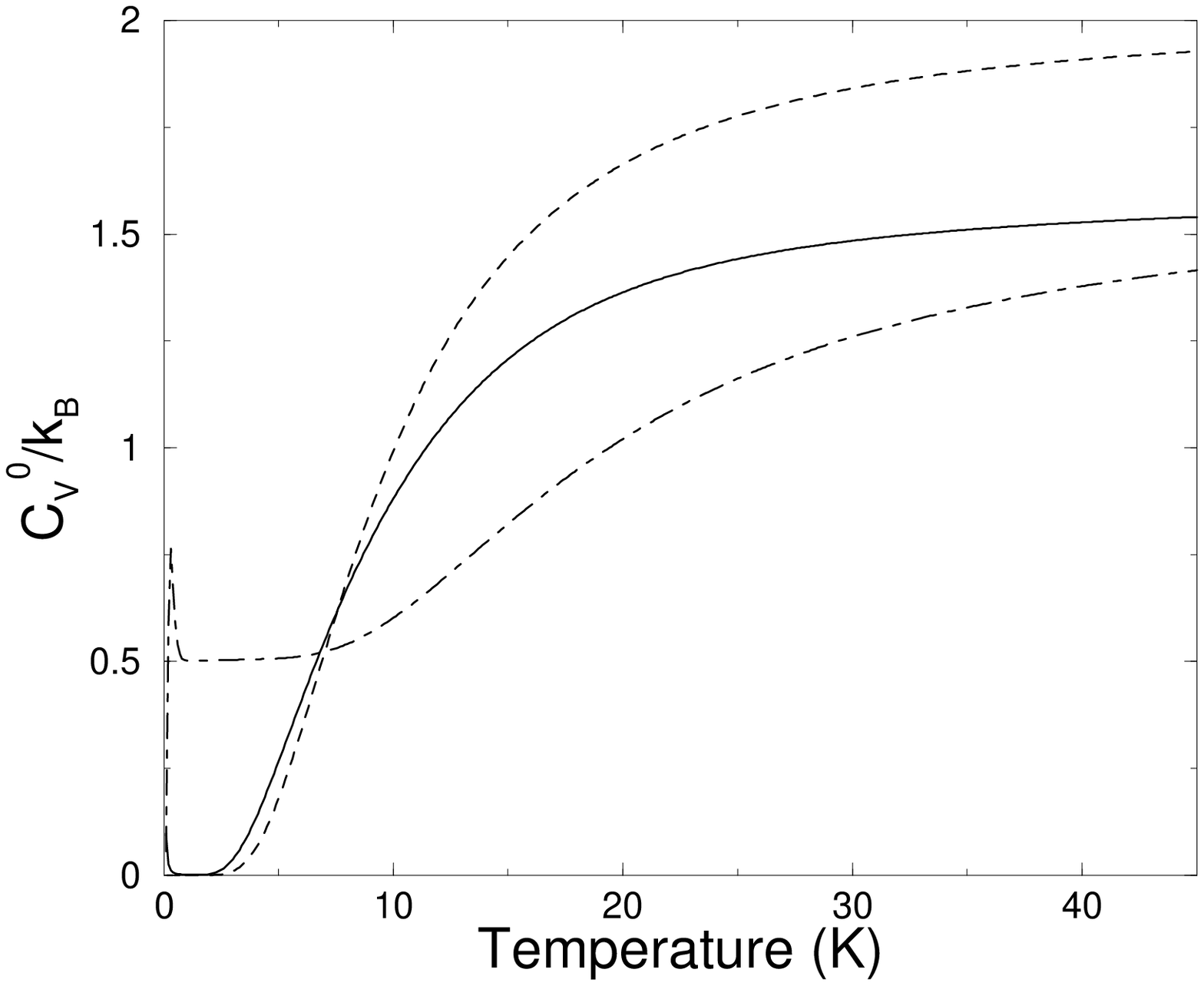}
  \end{minipage}
  \begin{minipage}{2.cm}
    \includegraphics[width=.9\textwidth]{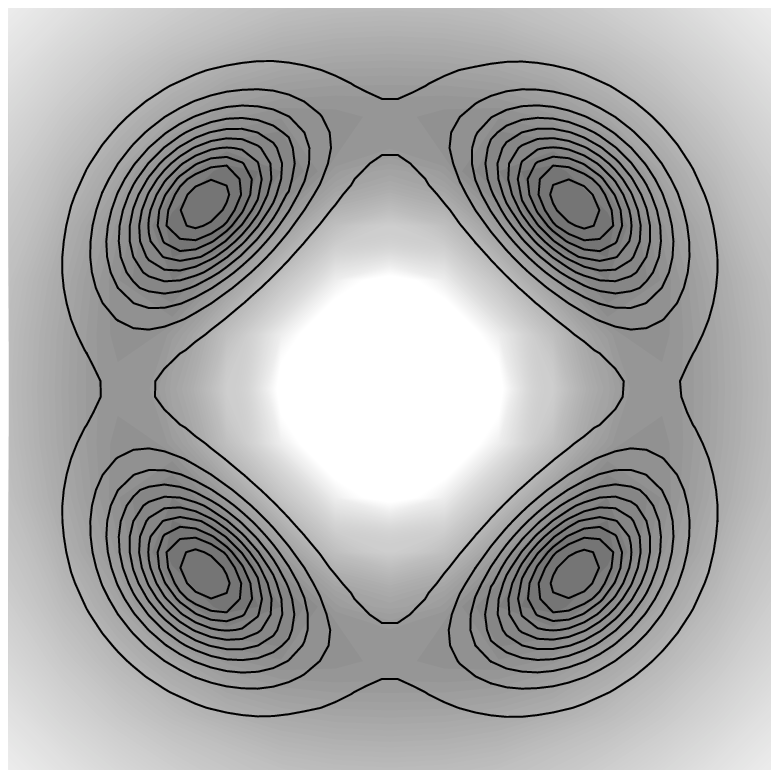}\\
    \includegraphics[width=.9\textwidth]{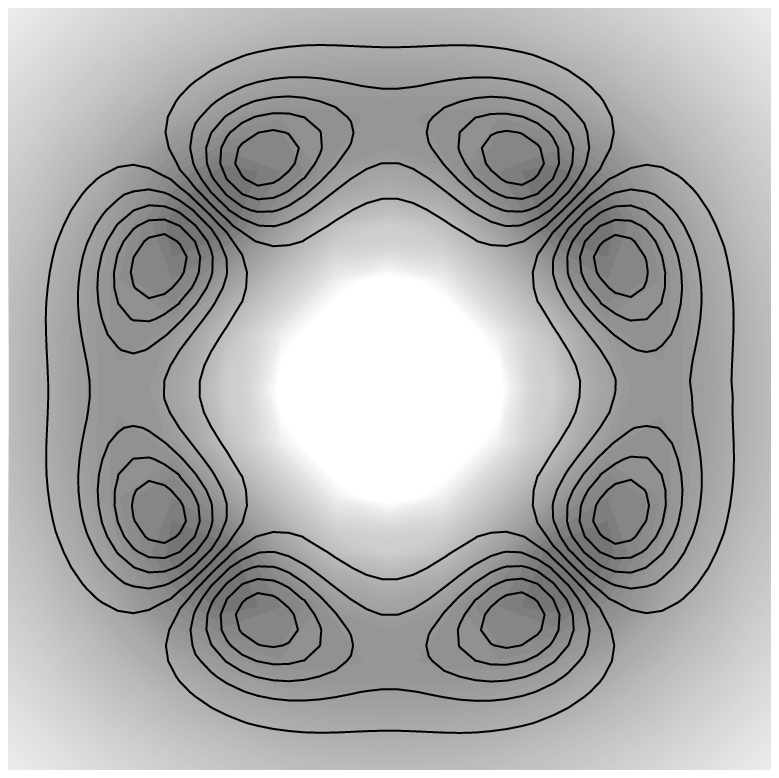}
  \end{minipage}
  \begin{center}
    \includegraphics[width=.35\textwidth]{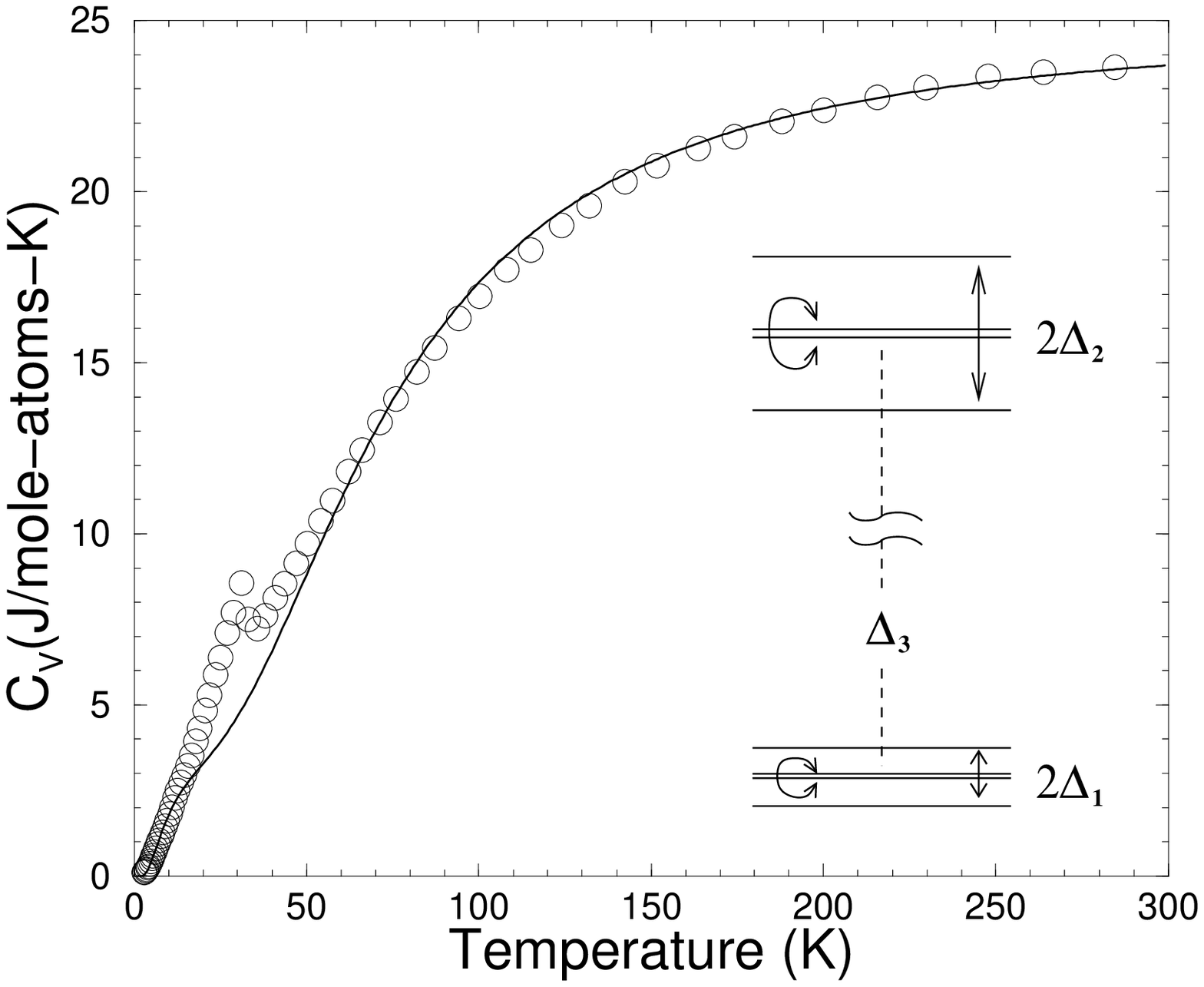}
  \end{center}
    \caption{\textit{Top left}: specific
      heat contribution C$_V^0$ of a guest atom. Dashed line:
      unperturbed 2D-HO, dash-dotted 
      line: 2D free rotor with $\bar{V_{1}}=307$ and $\bar{V_{0}}=0$,
      solid line: FLS's with $\bar{V_{1}}=292$ and $\bar{V_{0}}=30$.
      These parameters are in accord with the case of Eu
      (Table~\ref{tpar}).
      \textit{Top right}: contour plots of the ground state wavefunctions for
      FLS$_1$ and FLS$_2$. The shaded background corresponds to the
      four well potential. \textit{Bottom}: measured specific heat
      (circles) of \Eu{} from Ref. \cite{Sales01}. Solid line is the
      calculated total specific heat C$_V$ including the sum of 6 FLS/cell,
      HO contributions and the Debye term. It is
      almost identical to the fit using the pure Einstein oscillator
       \cite{Sales01} for the guest atom Eu. The peak around 30~K is due to the 
      ferromagnetic phase transition.
      \textit{Inset}: schematic representation of FLS$_{1,2}$.
      Arrows indicate transitions within each FLS system due to the
      quadrupolar interaction.\label{spheat}}
\end{figure}

Here, $\rho_{0}$ is the split site radial distance from the center,
corresponding to the maximum of the $n=0$ radial wave function. The
tunneling frequency $\Delta_{2}$ of the upper
FLS$_{2}$ and its energy relative to the lower one, $\Delta_{3}$,
may be calculated using~(\ref{delta}). $\Delta_{3}$ is a measure
for the size of the azimuthal corrugation in the radial minimum. The
low lying energy levels for the case of Eu
 clathrate, are shown in Fig.~\ref{eigenmu}. The
parameters for the three clathrates are summarized in
Table~\ref{tpar}. We note that the nuclear density
profiles for Sr and Ba do not correspond to the FLS wave
functions presented in the right part of Fig.~\ref{spheat}.
Our new model of azimuthal four-well tunneling enforces a
re-interpretation of low temperature properties in Eu clathrate.

%%%%%%%%%%%%%%%%%%%%%%%%%%%%%%%%%%%%%%%%%%%%%%%%%%%%%%%%%%%%%%%%%%%%%%%%%%%%
\begin{table}
\begin{tabular}{|c|c|c|c|c|c|c|c|}\hline
Atom &  $A$  &   $\rho_{0}/$~\r{A} & $\Theta/$K  & $\Delta_{1}/$K & 
$\Delta_{2}/$K& $\Delta_{3}/$K \\ \hline
Sr~\cite{Nolas00}  &  87.6 & 0.35  &   50       & 0.11 & 2.23 & 56 \\
Ba~\cite{Paschen01} & 137.3 & 0.20  &   70      & 2.19 & 4.45 & 110 \\
Eu~\cite{Sales01,Paschen01,Nolas00} & 152.0 & 0.43 & 30 & 0.05 & 0.85 & 22 \\
\hline
\end{tabular}
\caption{The parameters of "guest" atoms for the three
  clathrates. $A$ is the atomic mass number. $\Delta_{1}$,
  $\Delta_{2}$, and $\Delta_{3}$ are the 
  tunneling energies of the FLS$_{1}$, FLS$_{2}$, and the
  gap between the two, respectively, obtained for
  $\bar{V}_0=30$. $\rho_0$ for Eu is the average from the three
  references. (In Ref.~\cite{Paschen01} the position for Ba(2) is
  incorrect. Here the correct $\rho_0$ is used.)
  \label{tpar}}
\end{table}
%%%%%%%%%%%%%%%%%%%%%%%%%%%%%%%%%%%%%%%%%%%%%%%%%%%%%%%%%%%%%%%%%%%%%%%%%%%%%%%

%%%%%%%%%%%%%%%%%%%%%%%%%%%%%%%%%%%%%%%%%%%%%%%%%%%%%%%%%%%%%%%%%%%%%%%%%%%%%%%
%%%%%%%%%%%%%%%%%%%%%%%%%%%%%%%%%%%%%%%%%%%%%%%%%%%%%%%%%%%%%%%%%%%%%%%%%%%%%%
%
Using the eigenvalues from~(\ref{lambda0}), the specific heat, C(T),
may be numerically calculated. For the free rotor ($\bar{V_0}$ = 0)
closely spaced levels (Fig.~\ref{eigenmu}) lead to a constant C(T) for
moderately low T (Eq.~\ref{lambda0} and Fig.~\ref{spheat}). Turning on the
four well term, $\bar{V_0}$, leads to the ``bunching'' of the levels into two
FLS's, with large energy gap $\Delta_{3}$ between them. Since
$\Delta_{3}\simeq\Theta$ for Eu potential parameters and $\Delta_{1}$
is very small, 
C(T) for the pure HO and FLS model deviate very little. They are equally
acceptable, in contrast to the conventional TLS model. In addition, the
contribution of guest atoms to the total C(T) at high temperatures is
only around 10\%, so that the deviations in the high temperature
region are negligible, as seen in Fig.~\ref{spheat}.

Now we show that our new FLS model in addition leads to
characteristic elastic constant depressions observed in \Eu{} by RUS
measurements which the pure HO model cannot explain.
Assuming a quadrupolar interaction  $V\cos(2\phi)$ of FLS's
with the lattice, the matrix
elements between the eigenstates of Eq.~(\ref{mu}) can be
numerically calculated. The interaction matrix is nearly of
block diagonal form with negligible off-diagonal terms,
which is represented as direct sum of the two FLS's:
\begin{eqnarray}
\label{vint}
V_{int}=
\left(\begin{array}{rr}
\gamma_{1} X & 0 \\
0  &  \gamma_{2} X \\
\end{array}\right),
\end{eqnarray}
with $X=\sigma^x\otimes\sigma^x$ ($\sigma^x$ = Pauli matrix) and
$\gamma_1$, $\gamma_2$ the quadrupolar coupling strengths for
FLS$_1$ and FLS$_2$, respectively ($\gamma_1\ge\gamma_2$). The
possible transitions are shown in the inset of Fig.~\ref{spheat}.
The interaction term in Eq.~(\ref{vint}) is coupled to the elastic
strain $e_\mu$ leading to the total Hamiltonian
\begin{eqnarray}
H = \frac{1}{2}Vc_\mu^0e_\mu^2+e_\mu V_{int}+H_{FLS}.
\end{eqnarray}
Here, $c_\mu^0$ is the background elastic constant, $V$ is the 
sample volume and
$H_{FLS}$ the noninteracting two-FLS Hamiltonian.
The renormalised elastic constant is then given by the
static quadrupolar susceptibility of the FLS's according to
$\Delta c_{\mu} = (N_c/V_0)\chi_{Q}$ where $N_c=6$ is the number 
of Eu(2) atoms per unit cell of volume $V_0$ and
\begin{eqnarray}
\chi_Q&=&-\sum_{\lambda_{0,m}\neq\lambda_{0,n}}
\frac{|\langle m |V_{int}| n \rangle|^2}
{\lambda_{0,n}-\lambda_{0,m}}(p_m-p_n)\nonumber\\
&&-\frac{1}{k_B T}\{\sum_{\lambda_{0,m}=\lambda_{0,n}}
|\langle m |V_{int}| n \rangle|^2 p_m - \langle V_{int} \rangle^2\}
\end{eqnarray}
Here $p_m$ is the occupation factor for the FLS$_{1,2}$
tunneling state $|m\rangle$. The first (Van Vleck)
contribution is due to inelastic transitions within FLS$_{1}$
and FLS$_{2}$. The second (Curie) term is due to
elastic transitions within the degenerate doublets of both FLS$_{1}$
and FLS$_{2}$. 

The FLS's interact with the phonon bath. 
For a sufficiently strong interaction compared to the tunneling
energy the incoherent regime is entered. The crossover
temperature is given by
$T^{\ast}=\frac{1}{k_{B}}\sqrt{\frac{2\hbar}{\pi}\sqrt{\rho
v^{5}\hbar}\frac{2\Delta}{\gamma}}$~\cite{Wuerger96}.
$\Delta_{1}$ is an order of magnitude smaller than $\Delta_{2}$
(Table~\ref{tpar}) and correspondingly, T$^{\ast}$ for the
FLS$_{1}$ is significantly lower. In the coherent regime T$<$
T$^{\ast}$ the change of $\chi_Q$ is due to the relative change of
potential minima between the four wells. 
Because $\Delta_{1}$ is small its coherent tunneling
states are destroyed  by the coupling to the phonon bath already
at T $\sim$ 1 K. Then the guest atom may be considered as
statistically distributed over the four wells reducing the elastic
response to that of four independent wells. This is zero if only the
constant shift of a single well is considered. In contrast, for Eu parameters
FLS$_{2}$ is on the boundary of the coherent regime in the temperature
range of the depression. Therefore only FLS$_2$ contributes to the
quadrupolar response, explicitly:
\begin{eqnarray}
\label{dc}
\Delta c_\mu= -\frac{N_c}{V_0Z(T)}\,
e^{-\frac{\Delta_3}{k_BT}}
\left[\frac{2\gamma_{\mu 2}^2}{\Delta_2}
\sinh\left(\frac{\Delta_2}{k_BT}\right)
+\frac{2\gamma_{\mu 2}^2}{k_BT}\right]
\end{eqnarray}
%%%%%%%%%%%%%%%%%%%%%%%%%%%%%%%%%%%%%%%%%%%%%%%%%%%%%%%%%%%%%%%%%%%%%%%%%%%%%
\begin{figure}
    \includegraphics[width=.45\textwidth]{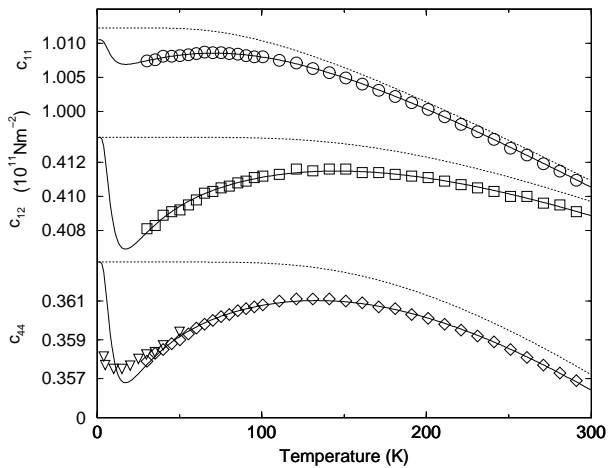}
    \caption{The fit of the elastic constants of Eu
      clathrate including 
      interaction with FLS$_{2}$ (solid thick line). Background
      variation corresponds to Varshni function~\cite{Varshni70}
      (dotted line). The depression with respect to the Varshni
      background is due to quadrupolar coupling to the upper FLS$_2$.
      Note the different scales for $c_\mu$.
      \label{elastic}}
\end{figure}
%%%%%%%%%%%%%%%%%%%%%%%%%%%%%%%%%%%%%%%%%%%%%%%%%%%%%%%%%%%%%%%%%%%%%%%%%%%%%
The depression in c$_\mu$ is caused by the thermal depopulation
of the relaxational FLS$_{2}$, described by the exponential factor in front of 
Eq.~(\ref{dc}). Within a conventional TLS model it may only be
described by assuming an unreasonably large tunneling frequency,
incompatible with the specific heat data, and unequal equlibrium well
depths which in the present context is unphysical because it violets
the fourfold symmetry.

%%%%%%%%%%%%%%%%%%%%%%%%%%%%%%%%%%%%%%%%%%%%%%%%%%%%%%%%%%%%%%%%%
\begin{figure}[b]
    \includegraphics[width=.45\textwidth]{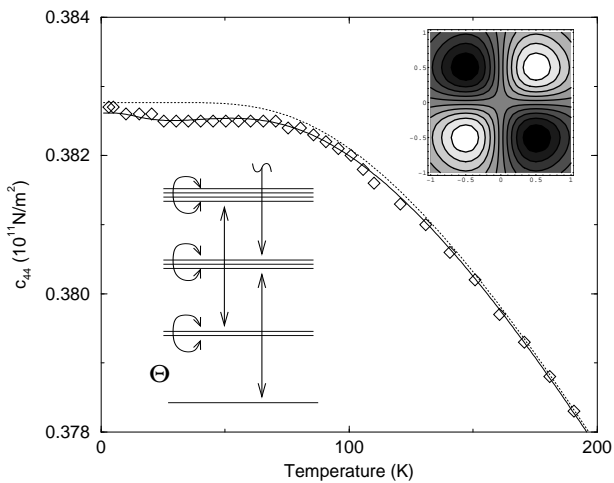}
    \caption{Elastic constant c$_{44}$ in \Sr. Fitted curve (solid
      thick line) is
      obtained using the Varshni function for the background (dotted
      line) and the 
      model of the  quadrupole interaction of the local Einstein mode
      with the elastic strain explained in the text. The interaction
      constant $\gamma_{HO}=0.015$~eV.\label{Sr}
      \textit{Inset}: schematic representation of the transitions
      between the states of 2D-HO due to the quadrupolar
      interaction. \textit{Upper inset}: the contour plot of the
      quadrupolar interaction potential 
      V$^{HO}_{int}$=$\gamma_{HO}\rho^2\sin(2\phi)
      \exp{[-\rho^2/(2a_0^2)]}$, $a_0=\hbar/\sqrt{K\,M}$.
      }
\end{figure}
%%%%%%%%%%%%%%%%%%%%%%%%%%%%%%%%%%%%%%%%%%%%%%%%%%%%%%%%%%%%%%%%%

The results for the Eu clathrate are shown in Fig.~\ref{elastic}.
Our fit determines $\gamma_{\mu 2}\approx0.073$~eV, with
$\Delta_{3}/k_{B}=22$~K. This value is used above to estimate
T$^{\ast}$ and to justify the neglect of the
FLS$_{1}$ contribution. The depression of c$_\mu$ is a direct
signature of the quadrupolar coupling to relaxational processes within
the upper FLS$_2$. Contrary to TLS model our FLS tunneling
model explains both
specific heat and elastic constant anomalies in \Eu{} without
contradiction and is consistent with the nuclear density profile
from the neutron diffraction. 
This presents strong evidence that a new type of FLS
tunneling states due to azimuthal tunneling of guest atoms has been
discovered in Eu clathrate.

In the case of \Sr{}  the split sites are less distinct than in \Eu,
with much of the nuclear density in the center~\cite{Sales01}, and
the four-well potential presented here is not appropriate. Due to the
large overlap, as may be deduced from the nuclear density profile,
the tunneling frequency may be very large. The split-site model
actually has the same agreement with the data as the single site
model~\cite{Chakoumakos00}. Therefore, we propose a different
model for \Sr. We assume that the Sr atoms sit in the shallow HO
potential that produces a soft Einstein mode. However we now propose a
quadrupolar form of interaction which rapidly decays at larger
distances from the center.
%At very large
%distances it has to increase again, but because of the oversized
%cages, we may assume this an off-center maxima.  
This is a crucial condition for a small dip in the elastic
constant to appear due to transitions between the excited
degenerate states of 2D-HO. In the upper inset of Fig.~\ref{Sr} this
interaction potential is shown. Although the four well guest atom
potential for Sr has not yet developed as in the Eu case, a tendency
is present in the model quadrupolar interaction potential which leads
to a satisfactory agreement with RUS data in Fig.~\ref{Sr}.

%Our elastic constant and specific heat measurements in Eu- clathrate have
%given direct evidence for the existence of a new type 
%tunneling states consisting of two partly degenerate four- level
%schemes in agreement with the observed nuclear density. In contrast
%the Sr- clathrate shows evidence of elastic coupling to doubly
%degenerate oscillator states. 

The RUS measurements were sponsored by NSF grant
DMR0206625. Oak Ridge National Laboratory is managed by
UT-Battelle, LLC, for the US Department of Energy under contract
DE-AC05-00OR22725.

\end{document}